\documentclass[letter,oldversion]{aa} 
\usepackage{graphicx}
\usepackage{txfonts}
\usepackage{natbib}
\bibpunct{(}{)}{;}{a}{}{,}

\begin{document}
 
\title{Rapid magnetic flux variability on the flare star CN~Leonis\thanks{Based on observations 
    collected at the European Southern Observatory, Paranal, Chile, 077.D-0011}}

\author{A. Reiners
  \inst{1,2}\fnmsep\thanks{Marie-Curie International Outgoing Fellow}
  \and
  J. H. M. M. Schmitt\inst{1}
  \and
  C. Liefke\inst{1}
}

\offprints{A. Reiners}

\institute{            Hamburger Sternwarte, Gojenbergsweg 112, D-21029 Hamburg, Germany\\
  \email{jschmitt, cliefke@hs.uni-hamburg.de}
  \and
  Universit\"at G\"ottingen, Institut f\"ur Astrophysik, Friedrich-Hund-Platz 1, D-37077 G\"ottingen, Germany\\
  \email{Ansgar.Reiners@phys.uni-goettingen.de}
}

\date{Received 15 January 2007 / Accepted 28 February 2007}


\abstract {We present UVES/VLT observations of the nearby flare star
  CN~Leo covering the Wing-Ford FeH band near 1~$\mu$m with high
  spectral resolution.  Some of the FeH absorption lines in this band
  are magnetically sensitive and allow a measurement of the mean
  magnetic flux on CN~Leo.  Our observations, covering three nights
  separated by 48 hours each, allow a clear detection of a mean
  magnetic field of $Bf \approx 2.2$\,kG. The differential flux
  measurements show a night-to-night variability with extremely high
  significance. Finally, our data strongly suggest magnetic flux
  variability on time scales as low as 6 hours in line with
  chromospheric variability.}


\maketitle
%

\section{Introduction}

Stellar activity can be observed in various tracers from all layers of
a stellar atmosphere: Doppler imaging techniques reveal spots in the
photosphere and trace regions of large magnetic flux, chromospheric
emission shows up in specific lines like e.\,g. H$\alpha$ or
\ion{Ca}{ii}~H~\&~K originating above photospherically active regions,
and finally, X-rays trace the hottest material and allow the
investigation and characterization of coronal properties.  In active
stars, these activity indicators can exceed the levels observed on the
Sun by several orders of magnitude.

The strong activity observed in many stars implies strong magnetic
fields, which are thought to lie at the root of all observed activity
phenomena. Yet the direct measurement of magnetic fields in stellar
photospheres is difficult. Most of the methods to measure magnetic
fields require the detection of Zeeman broadening in magnetically
sensitive spectral lines. Magnetic field strengths up to several
kilo-Gauss and large filling factors have been measured for a few
early M~dwarfs and T~Tauri stars from optical \ion{Fe}{i} and infrared
\ion{Ti}{i} lines \citep[for an overview see][]{JKV00}.
Spectropolarimetry allows to reconstruct the field geometry for some
of these objects \citep{Donati_spectropolarimetry}.  However,
appropriate atomic lines for such measurements either disappear or
become more and more blended with molecular lines for objects later
than spectral type mid M.  \citet{Reiners06} introduced a method to
measure the mean magnetic flux of these ultracool dwarfs from lines of
the FeH absorption band around 9950~\AA. The FeH band comprises a
large number of absorption lines with very different Zeeman
sensitivity. Lines with small sensitivity can be used to determine the
rotation velocity, and magnetically sensitive lines unambiguously
indicate the quantity magnetic field strength~$\times$~filling factor
($Bf$).  \citet{Reiners07} successfully measured the mean magnetic
field in a sample of 22 M~dwarfs with different activity levels and
demonstrated that the magnetic flux of their sample objects is
strongly correlated with H$\alpha$ emission.

Stellar activity is inherently variable on time scales from seconds to
decades.  Variability on the shortest time scales occurs in the form
of sudden outbursts or flares, which can be observed in several
wavelength bands.  While this variability has been extensively studied
through proxy indicators like those mentioned above, we tackled the
challenge to directly measure magnetic field changes and correlate the
results with proxies like, e.\,g., H$\alpha$ emission.  For this
purpose, we chose to study the flare star \object{CN~Leo} (Gl\,406).
At a distance of 2.39\,pc \citep{Henry}, CN~Leo is one of the nearest
late-type stars and also among the objects examined for magnetic
fields by \citet{Reiners07}, who estimated an integrated surface
magnetic flux of $Bf \approx 2.4$\,kG. The temperature of CN~Leo is
2800~K to 2900~K \citep{Pavlenko, Fuhrmeister05} at a spectral type of
M5.5 or M6.0 \citep{Reid,Kirkpatrick91}.  Further information in
particular on the coronal properties of CN~Leo can be found in
\citet{Fuhrmeister07}. CN~Leo has -- fortuitously but somewhat
surprisingly -- a rather low apparent rotation velocity of $v \sin i
\approx 3.0$\,km\,s$^{-1}$ \citep{Mohanty,Reiners07}, thus making it
an ideal target for spectroscopic high-resolution studies of stellar
magnetic activity.

\section{Observations and data reduction}

CN~Leo was observed during three half nights of May 19/20, 21/22 and
May 23/24, 2006 with UVES/VLT.  The UVES spectrograph was operated in
dichroic mode with spectral coverage from 3200~\AA\ to 3860~\AA\ in
the UV and from 6400\,\AA\ to 10\,080\,\AA\ with a gap between
8190\,\AA\ and 8400\,\AA\ in the red. We used a slit width of 1$''$
resulting in a resolving power of $R \sim 40\,000$.  Here we consider
only the far red part of our spectra on the upper chip in the red arm.
Exposure times varied according to seeing conditions from 100\,s to
300\,s, resulting in a total of 68, 24 and 89 spectra taken in our
three half nights.  The data were processed using the IDL based {\tt
  REDUCE} reduction package \citep{reduce}. Since the MIT/LL CCD-20
chip shows very strong fringing in the red wavelength region, we
performed a 2D-flatfielding with standard calibration lamp exposures.
The typical signal-to-noise ratio (SNR) of a single spectrum is on the
order of 40 around 1~$\mu$m.

\section{Method of Analysis}

In order to determine the mean magnetic field strength in CN~Leo we
use the molecular FeH band near 1\,$\mu$m; this band contains a set of
narrow, individual absorption lines, some of which are magnetically
very sensitive, while others are not. Unfortunately, so far no Land\'e
g-factors are available for these lines.  Therefore, magnetic flux
measurements are performed by comparing FeH lines to spectra of stars
with known magnetic flux and known activity, i.e., to template stars.
The feasibility and performance of this method has been successfully
demonstrated by \cite{Reiners07}.  As a second measurement of the
magnetic flux, we use the line ratio method suggested by
\cite{Reiners06} as a consistency check after we calculated the
magnetic flux with the line-profile fitting method.

\subsection{Line profile fitting}

\begin{figure}
  \resizebox{\hsize}{!}{\includegraphics{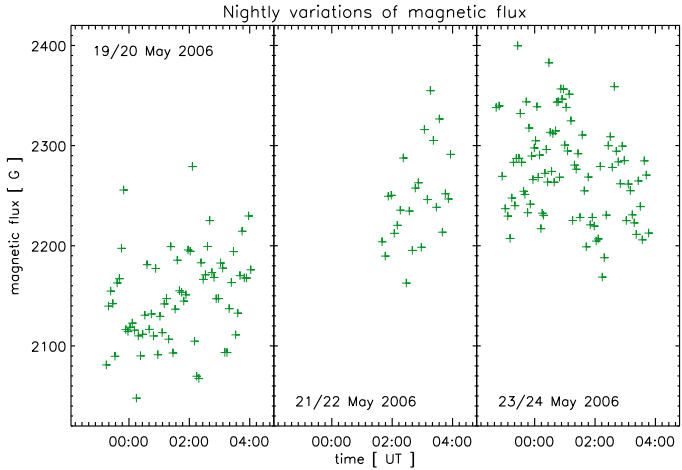}}
  \caption{Variation of the mean magnetic flux $Bf$ over the three
    nights. The relative error on a an individual data point is about
    80 G (see text).}
  \label{fig:Bf_Time}
\end{figure}

To determine the mean magnetic field of CN~Leo we follow the approach
by \citet{Reiners07}, who analyzed HIRES Keck data of 22 ultracool
dwarfs.  The basic idea of this method consists in modeling the
observed spectrum of the target star as a linear combination of the
spectrum of a magnetically inactive star with very low magnetic flux
(we chose GJ~1002, M5.5) and that of a magnetically very active star
(we chose Gl~873, M3.5) with $Bf \approx 3.9$\,kG \citep{JKV00}. As
actual templates we used the Keck HIRES spectra of Gl~873 and GJ~1002.
A comparison of the UVES and Keck spectra of CN~Leo shows that the
resolving power of the HIRES and UVES spectra are very much alike. We
therefore employed both data sets without any further assumptions on
the instrument profiles.

Specifically, we construct a linear interpolation of the
normalized spectrum $S_{\rm CN\,Leo}$ of CN~Leo from the normalized 
template spectra $S_{\rm Gl\,873}$ and $S_{\rm Gl\,1002}$ through
\begin{equation}
\label{eq:S}
S_{\rm CN\,Leo} = pS_{\rm Gl\,873} + (1-p) S_{\rm GJ\,1002}.
\end{equation}

The weight $p$, with $0 < p < 1$, yields $Bf = p~Bf({\rm Gl\,873}) =
p~3.9$\,kG. We find the best value of $p$ by $\chi^2$-minimization
\citep{NR} through a comparison of $S_{\rm CN\,Leo}$ with the actually
measured (normalized) spectrum.  Note that a few steps are required
before the interpolation of the two template spectra can be computed:
Both template spectra are adjusted to match the target spectrum in
line depth and in the overall width of the magnetically insensitive
lines \citep[for more details, see][]{Reiners07}. The widths of the
magnetically insensitive lines are used to calibrate the line
broadening that affects all spectral lines. This global broadening is
dominated by stellar rotation and the spectrograph's resolving power.
While stellar rotation can obviously be assumed constant during all
observations, the spectral resolution can also be affected by the
seeing conditions during observation. In a perfect spectrum, for a
star with low rotation velocities this effect would mimic the effect
of extra rotation and should show up if we formally allow the fit to
also adjust the projected rotation velocity $v\,\sin{i}$ for each
spectrum.

We performed the $\chi^2$-minimization first treating $v\,\sin{i}$,
$Bf$ (i.e., $p$, see Eq.\,\ref{eq:S}), and the line depth $a$ as free
parameters. However, we did not find any correlation between
$v\,\sin{i}$ and the seeing conditions that we measured through the
widths of telluric lines. We therefore believe that small extra
broadening due to variable seeing conditions does not significantly
affect the line widths of the absorption lines, which are dominated by
temperature and stellar rotation.  Treating $v\,\sin{i}$ as a free
parameter we found $v\,\sin{i}$ values between 2.0 and
2.5\,km\,s$^{-1}$, which are only marginally different given the
comparably low resolution (1 resolution element $\approx
7.8$\,km\,s$^{-1}$).\footnote{Note that $v\,\sin{i}$ is the broadening
  required to match the template spectra and the spectra of CN\,Leo,
  thus it is only a lower limit for the true projected rotation
  velocity of CN\,Leo.}  Since the rotation velocity of CN\,Leo is not
variable on the timescales of our measurements, we took the mean value
of all values of $v\,\sin{i}$ from the fit, $\overline{v\,\sin{i}} =
2.14$\,km\,s$^{-1}$ (incidentally, the value of the mean coincides
with the value of the median, so it is not necessary to decide which
measure is best here).  We then repeated the fit, now with a fixed
value of $v\,\sin{i} = 2.14$\,km\,s$^{-1}$. This procedure did not
significantly change the derived results when compared to the results
achieved with $v\,\sin{i}$ as a free parameter. In particular, the
derived $Bf$ values do not vary significantly, and we proceed using
the results achieved with the fixed value of $v\,\sin{i}$.

A total of 181 spectra of CN~Leo were recorded during our campaign and
were used to derive mean magnetic fields as described above.  In
Fig.\,\ref{fig:Bf_Time}, we plot all our $Bf$ measurements versus
time.  The typical uncertainty of our $Bf$ measurements is
approximately $\overline{\Delta Bf} = 80$\,G; this value represents
the 1-$\sigma$ uncertainty from the $\chi^2$-fit, i.e., the values of
$Bf$ for which $\chi^2 < \chi^2_\mathrm{min} + 1$ \citep{NR}. We
emphasize that this error is of purely statistical nature and cannot
be identified as the true uncertainty of the \emph{absolute}
measurement of the magnetic flux $Bf$. For the \emph{relative}
measurement, however, this is the uncertainty that we adopt. We note
that our fitting procedure started with an initial guess of $Bf =
2500$\,G for every spectrum, and that the algorithm did significantly
modify this initial guess. We also tried the effect of a different but
fixed value of $v\,\sin{i}$ and found that a different $v\,\sin{i}$
slightly modifies the absolute values of all measurements, but does
not change our results on the differential measurements.

\subsection{Analysis of spectral differences}

\begin{table}[h]
  \centering
  \caption{\label{tab:nights}Mean magnetic flux, standard deviation, error of the nightly mean,  
    number of observations per night for all three nights; the other columns contain results of non-parametric statistical tests (see text).}
  \begin{tabular}{ccccccc}
    \hline
    \hline
    \noalign{\smallskip}
    Night & $\overline{Bf} \pm \sigma$ [G]& $N$ & KS1 & KS2 & RS1 & RS2 \\
    \noalign{\smallskip}
    \hline
    \noalign{\smallskip}
    1 & 2148 $\pm$ 45 $\pm$ 5.5 & 68 & --           & --              & -- & --\\
    2 & 2248 $\pm$ 47 $\pm$ 9.6 & 24 & $7\,10^{-11}$& --              & 0 & --\\
    3 & 2275 $\pm$ 48 $\pm$ 5.1 & 89 & $3\,10^{-28}$& $ 8.4\,10^{-3}$ & 0 & $ 7.7\,10^{-3}$\\
    \noalign{\smallskip}
    \hline
  \end{tabular}
\end{table}

Is there any true variation of $Bf$ visible in our data? The
difference in the magnetic flux between the individual spectra with
the maximum and the minimum flux is $\approx$ 350\,G, i.e., less than
10\,\% of the difference in flux between the two template stars of
$\approx$ 3900\,G.  At the spectral type of CN~Leo, the line depth of
the magnetically most sensitive lines varies by $\approx$ 15\,\% of
the continuum value \citep[e.g. the line at 9948\,\AA\ varies between
$\sim 0.58$ and $\sim 0.74$, cp. Fig.\,6 in][]{Reiners07}. Thus, in
order to detect a magnetic flux variation of 350\,G, a variation on
the order of 1.5\,\% has to be detected in this line, i.e., a SNR on
the order of 70 is required. Because of the short individual exposures
such a SNR is not available in our spectra, and therefore a direct
comparison cannot be expected to yield a significant difference in the
Zeeman signatures directly visible in the spectra.  However, our
$\chi^2$-fit uses simultaneously the information contained in many
lines and hence is superior to a direct comparison of line depths.

Nonetheless, our $\chi^2$-fits could be fooled by a variety of effects
that are difficult to control if the results cannot be independently
verified.  In order to achieve such a consistency check, we
constructed two mean spectra by averaging all spectra from the first
night (68) and all spectra from the third night (89). The mean values
of $Bf$ for the three different nights are given in
Table\,\ref{tab:nights}. Between the first and the third night
$\overline{Bf}$ differs by 125\,G. The summation of all spectra in the
first and third night with a SNR of typically $\sim$ 40 each yields
average spectra with a SNR on the order of $\sim$ 300.  Since a SNR on
the order of 200 is required to detect a difference of 125\,G in the
9948\,\AA\ lines between the two mean spectra (see above), the
difference in the magnetically most sensitive lines should become
visible in such a comparison.

\begin{figure}
  \centering
  \mbox{\includegraphics[width=.5\textwidth]{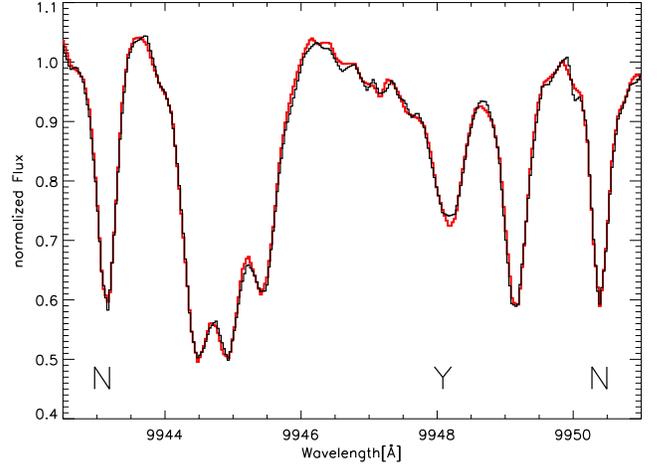}}
  \caption{Co-added spectra from the 68 spectra of the first night
    (red) and from the 89 spectra of the third night (black).  At the
    achieved SNR $\approx$ 300 Zeeman broadening becomes visible in
    the most sensitive lines. Y: magnetically sensitive line, N:
    magnetically insensitive lines.}
  \label{fig:compare}
\end{figure}

In Fig.\,\ref{fig:compare}, we show the mean spectra for the first and
third nights. The magnetically very sensitive line at 9948\,\AA\ is
marked with a ``Y'', two relatively insensitive lines are marked with
an ``N''. As expected from our estimate, a small difference becomes
visible in the magnetically sensitive line, while insensitive lines do
not show systematic variation.

\subsection{Magnetically sensitive line ratios}

As a second consistency check we use the two mean spectra for the
first and third nights and calculate the two line depth ratios as
suggested by \citet{Reiners06}. Each ratio is calculated from
measuring the depth of two neighbored lines, one of which is Zeeman
sensitive and one is insensitive. We use the two ratios
$9949.11/9948.13$ and $9956.77/9958.25$ as in \cite{Reiners06}, both
ratios should be larger at high magnetic flux. The results are given
in Table\,\ref{tab:ratios} and should be compared to Fig.\,7 of
\cite{Reiners06}. That figure shows that indeed the $9949.11/9948.13$
ratio is more sensitive to magnetic flux variation than the
$9956.77/9958.25$ ratio is. Both ratios and their variation are in
good agreement with our results from $\chi^2$-fitting.

\begin{table}[h]
  \centering
\caption{\label{tab:ratios}Line depth ratios as introduced in \cite{Reiners06}. 
Both ratios are consistent with our results (see text).}
  \begin{tabular}{lcc}
    \hline
    \hline
    \noalign{\smallskip}
    & 9949.11/9948.13 & 9956.77/9958.25 \\
    \noalign{\smallskip}
    \hline
    \noalign{\smallskip}
    $Bf_\mathrm{Night 1}$ & $1.48 \pm 0.01$ & $1.44 \pm 0.01$\\
    $Bf_\mathrm{Night 3}$ & $1.59 \pm 0.01$ & $1.45 \pm 0.01$\\
    \noalign{\smallskip}
    \hline
  \end{tabular}
\end{table}

\section{Results, Discussion and Conclusions}

As a first result of our analysis of 181 spectra of CN~Leo we note
that magnetic flux is consistently detected in each UVES spectrum.
At first sight the data are consistent with the assumption of a
constant magnetic flux during each night, while the magnetic flux does
appear to vary from night to night.  In Table\,\ref{tab:nights}, we
give the derived mean values of $Bf$ and their standard deviation for
all three nights. The standard deviation is consistent with the
estimated uncertainty of an individual observation; in fact, it is a
even little smaller, which could mean that the differential
measurements are somewhat more precise than estimated.

The uncertainty in the absolute value of the field is dominated by our
ignorance of the splitting patterns of the magnetically sensitive FeH
lines. Our method only employs a linear interpolation between the two
spectra of the template stars, i.e., it assumes that the splitting
patterns are the same ($B$ is constant), and that only the filling
factor $f$ is variable. This is probably not the case, although the
change in filling factor is usually assumed to be the dominant factor.
At any rate, we cannot yet calculate different splitting patterns and
thus, following \cite{Reiners07}, we estimate the uncertainty on the
absolute value to be on the order of a few hundred Gauss.  However, here we
are interested in the accuracy of the absolute measurements only to a
lesser extent, rather are we interested in the \emph{differential}
measurements comparing the values of $Bf$ during several nights and
within a given night.  Since the magnetically insensitive lines stay
constant, they provide a very accurate calibration, particularly when
intercomparing the exposures. Thus, we estimate that our uncertainties
in the individual differential measurements are indeed on the order of
the uncertainty of the $\chi^2$-fit, i.e., 80\,G.

Inspecting the nightly mean magnetic field values and their dispersion
(Table \ref{tab:nights}), the difference in these means appears to be
significant only on the 1-2 $\sigma$ level when compared to the
variance of the distribution.  We checked the nature of the
distribution of our nightly measurements around the respective nightly
means and found very good agreement with a Gaussian distribution and
therefore the accuracy of the mean ought to be much smaller than the
width of the observed distribution.  Computing the uncertainty of the
nightly means, we find values reduced by $\sqrt{N}$ (cf.,
Tab.\ref{tab:nights}) and the differences between the individual
become very significant indeed.  We verified the existence of
night-to-night magnetic field variations also by analyzing the
actually measured distributions of $Bf$ using two non-parametric test
methods. We specifically applied the Wilcoxon rank sum test (denoted
by RS in Table \ref{tab:nights}) and the two sample Kolmogorov-Smirnov
test (KS) to investigate whether the measured $Bf$ during our three
observing nights are drawn from the same parent population (KS1 and
RS1 indicate comparison to the first night, KS2 and RS2 to the
second). The derived probabilities from the KS- and RS-tests that the
null hypothesis "no night-to-night magnetic flux change" is correct
are also listed in Table \ref{tab:nights}.  As is clear from Table
\ref{tab:nights}, the field measurements for the nights 1 and 2 and 1
and 3 differ in an extremely significant fashion, and even the
measurements between nights 2 and 3 appear to differ at a confidence
level in excess of 0.99 independent of the test statistics used. In
summary, the above statistical analysis strongly suggests magnetic
flux changes on time scales of 48 hours.

\begin{figure}
  \centering
  \mbox{\includegraphics[width=.5\textwidth]{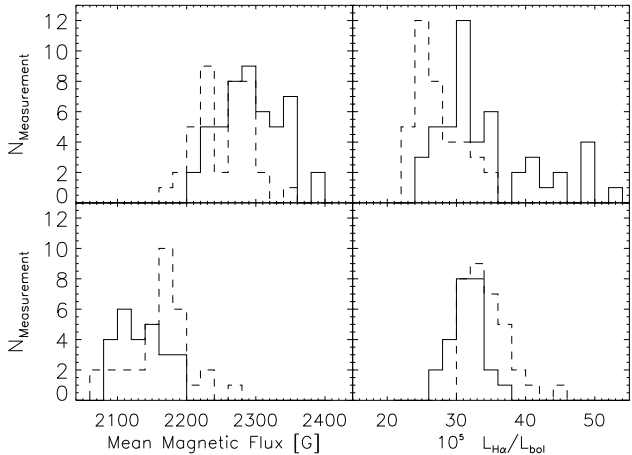}}
  \caption{Magnetic flux (left panels) and H$\alpha$ emission (right
    panels) for first night (lower panels) and third night (upper
    panels); in each panel, data for the respective (half-)nights are
    shown for the first half (solid histograms) and second half
    (dashed).  Note the correlated change in magnetic flux and
    H$\alpha$ emission during the first and third half night.  Eight
    data points in a large flare during the first half of the first
    night are not shown because of saturation of the H$\alpha$ line
    core.}
  \label{fig:short}
\end{figure}

How about the presence of flux changes on shorter time scales? In
order to address this issue we subdivided the available magnetic flux
measurements for each of our three nights into two halves and applied
the same tests performed for night-to-night variability to assess any
intra-night variability.  While we cannot claim any intra-night
variability during night 2 (with only 24 $Bf$ measurements available),
we find significant intra-night variability for night 1 and night 3.
The variability is such that the magnetic flux during night 1 appears
to be increasing, while it is decreasing during night 3.
Interestingly, the intra-night variability in chromospheric emission
lines follows the same pattern as the magnetic flux. We measured the
chromospheric flux from the H$\alpha$ lines contained in our spectra
\citep[see][]{Reiners07} and calculated
$L_{\rm{H}\alpha}/L_{\rm{bol}}$ with the same value of $L_{\rm{bol}}$
for all spectra. The change in magnetic flux and H$\alpha$ emission is
directly visible in the distribution of magnetic flux and H$\alpha$
emission during the first and second quarters of nights 1 and 3 (first
and second half of our observation in each night). The distribution
for nights 1 and 3 are shown in the lower and upper panels of
Fig.~\ref{fig:short}, respectively. We find the difference between the
two distributions shown in each panel of Fig.~\ref{fig:short} is
statistically highly significant.

These trends are verified by formally describing the nightly magnetic
flux measurements as a linear function with time.  Linear regression
analysis shows that including a linear trend significantly better
describes the data as opposed to the assumption of a constant magnetic
flux; the values of $\Delta \chi^2$ are 6.0 and 7.3 for nights 1 and 3
respectively.  We find a positive flux change during night 1 and a
negative change during night 3; the observed flux changes to an amount
of $\approx$ 50\,G over a time scale of 6~hours.  While we cannot
entirely exclude artifacts of the fit procedure as the cause for the
intra-night variations, two facts make us confident that the observed
changes are of a physical origin: First, the observed changes are of
opposite sign during night 1 and 3. Second, the magnetic flux changes
during those nights correlate with corresponding activity changes
during the same nights, which we have verified in the H$\alpha$ lines
contained in our spectra (see Fig.\ref{fig:short}). We note that our
spectra contain a multitude of chromospheric emission lines, which are
correlated with H$\alpha$ emission.  These will be analyzed in a
subsequent publication.

\citet{Reiners07} measured $Bf = 2.4$\,kG in a spectrum of CN~Leo
taken in 2005. This result is on the upper end of the values we found
during our observations. However, their spectra were not taken at the
same instrument, which introduces uncertainties due to the different
calibrations. Hence their result is consistent with our measurements.

In summary, we have successfully measured a mean magnetic field of $Bf
\approx 2.2$\,kG on the flare stars CN~Leo using the FeH bands near
1\,$\mu$m.  Our monitoring campaign shows definite magnetic flux
changes of $\sim 100$\,G on a time scale of 48~hours.  Further, the
data of the individual (half) nights suggest flux changes on even
shorter time scales, which are correlated with the general activity
level of CN~Leo as measured in H$\alpha$ emission.  Further monitoring
of magnetic fields and their variations in CN~Leo and similar stars
appears very promising.

\begin{acknowledgements}
  We thank Prof. G. Basri for letting us use his Keck spectra of
  GJ~1002 and Gl~873, and the referee, Christopher Johns-Krull for a
  very constructive report. AR has received research funding from the
  European Commission as an Outgoing International Fellow
  (MOIF-CT-2004-002544).  CL acknowledges financial support from DLR
  under 50OR0105.
\end{acknowledgements}


\begin{thebibliography}{}

\bibitem[{{Donati} {et~al.}(2006), {Donati}, {Forveille}, {Cameron},
    {Barnes}, {Delfosse}, {Jardine}, \&
    {Valenti}}]{Donati_spectropolarimetry} {Donati}, J.-F.,
  {Forveille}, T., {Cameron}, {et~al.} 2006, Science, 311, 633-635

\bibitem[{{Fuhrmeister} {et~al.}(2007){Fuhrmeister07}, {Schmitt},
  \& {Liefke}}]{Fuhrmeister07}
{Fuhrmeister}, B., Liefke, C., \& {Schmitt}, J.~H.~M.~M., 
  2007, \aap, submitted

\bibitem[{{Fuhrmeister} {et~al.}(2005){Fuhrmeister05}, {Schmitt},
  \& {Hauschildt}}]{Fuhrmeister05}
{Fuhrmeister}, B., {Schmitt}, J.~H.~M.~M., \& {Hauschildt}, P.~H.
  2005, \aap, 439, 1137

\bibitem[{{Henry} {et~al.}(2004){Henry}, {Subasavage}, {Brown}, {Beaulieu},
  {Jao}, \& {Hambly}}]{Henry}
{Henry}, T.~J., {Subasavage}, J.~P., {Brown}, M.~A., {et~al.} 2004, \aj, 128,
  2460

\bibitem[Johns-Krull \& Valenti, 2000]{JKV00}Johns-Krull, C., \&
  Valenti, J.A., 2000, ASPC, 198, 371

\bibitem[{{Kirkpatrick} {et~al.}(1991){Kirkpatrick}, {Henry}, \&
  {McCarthy}}]{Kirkpatrick91}
{Kirkpatrick}, J.~D., {Henry}, T.~J., \& {McCarthy}, D.~W. 1991, \apjs, 77, 417

\bibitem[{{Mohanty} \& {Basri}(2003)}]{Mohanty}
{Mohanty}, S. \& {Basri}, G. 2003, \apj, 583, 451

\bibitem[{{Pavlenko} {et~al.}(2006){Pavlenko}, {Jones}, {Lyubchik}, {Tennyson},
  \& {Pinfield}}]{Pavlenko}
{Pavlenko}, Y.~V., {Jones}, H.~R.~A., {Lyubchik}, Y., {Tennyson}, J., \&
  {Pinfield}, D.~J. 2006, \aap, 447, 709

\bibitem[{{Piskunov} \& {Valenti}(2002)}]{reduce}
{Piskunov}, N.~E. \& {Valenti}, J.~A. 2002, \aap, 385, 1095

\bibitem[Press et al., 1992]{NR} Press W.H., Teukolsky S.A.,
  Vetterling W.T., \& Flannery B.P., 1992, Numerical Recipes in C,
  Cambridge Universtity Press

\bibitem[{{Reid} {et~al.}(1995){Reid}, {Hawley}, \& {Gizis}}]{Reid}
{Reid}, I.~N., {Hawley}, S.~L., \& {Gizis}, J.~E. 1995, \aj, 110, 1838

\bibitem[Reiners \& Basri, 2006]{Reiners06}Reiners, A., \& Basri, G., 2006, 
  \apj, 644, 497

\bibitem[Reiners \& Basri, 2007]{Reiners07}Reiners, A., \& Basri, G., 2007, 
  \apj, 656, 1121

\end{thebibliography}
\end{document}